\documentclass[letter]{aa}

\usepackage{graphicx}
\usepackage{txfonts}
\usepackage{hyperref}
\usepackage{amsmath}
\usepackage{ulem}

\newcommand{\ximmlin}{\xi_{\mathrm{mm}}^{\mathrm{lin}}}
\newcommand{\dfq}{\Delta f_{\mathrm{Q}}}
\newcommand{\gn}{G_n}

\begin{document}

\title{Assembly bias from nuisance to probe}
\subtitle{I: the relation between galactic conformity and the linear matter clustering}
\titlerunning{The relation between galactic conformity and the linear matter clustering}

\author{
Nelson Padilla\inst{\ref{Iate},\ref{OAC}}\and
Ivan Lacerna\inst{\ref{uda}} \and
Dante Paz\inst{\ref{Iate},\ref{OAC}}
}  

\institute{
CONICET. Instituto de Astronomía Teórica y Experimental (IATE). Laprida 854, Córdoba X5000BGR, Argentina \label{Iate}\\
\email{nelson.padilla@unc.edu.ar}
\and 
Universidad Nacional de Córdoba (UNC). Observatorio Astronómico de Córdoba (OAC). Laprida 854, Córdoba X5000BGR, Argentina\label{OAC}
\and
Instituto de Astronom\'ia y Ciencias Planetarias, Universidad de Atacama, Copayapu 485, Copiap\'o, Chile \label{uda} 
}

\date{\today}

\abstract
{Two-halo galactic conformity is commonly interpreted as a manifestation of galaxy assembly bias, but its statistical structure and physical origin remain unclear.}
{We aim to write the quenched-neighbour statistic in correlation-function form, test whether its scale dependence follows the linear matter correlation function $\ximmlin(r)$, and separate the contributions of halo-mass bias and assembly bias to its amplitude.}
{Using galaxies in IllustrisTNG300-1 at $z=0$, we measure the two-halo galactic conformity statistic of quenched neighbours at distance $r$, $\dfq(r)$, and related quantities in real space, compute the required correlations, perform shuffling tests at fixed halo mass, compare several $\Delta f$ observables, and explore the transformed family $G_n$.}
{We show explicitly that $\dfq(r)$ can be written directly in terms of
correlation functions and that, over
$\sim 2$--$40\,h^{-1}\,\mathrm{Mpc}$, it is well described by
$A_{\rm fit}\,\ximmlin(r)$. {Thus nonlinear and baryonic terms do not
dominate the residual scale dependence isolated by this statistic.}
{Halo-mass bias alone predicts lower amplitudes than measured, while
fixed-mass shuffling strongly suppresses the signal; in TNG300 the amplitude is
therefore dominated by 
galaxy assembly bias at fixed halo mass.} Quenching,
colour, and concentration share a common rescaled shape, whereas stellar-mass
and halo-mass splits do not.}
{These results suggest that {galaxy} assembly bias sets the amplitude of two-halo conformity, while the double-difference structure of the statistic suppresses nonlinear residuals when the compared populations have similar halo-mass and transition-scale structure.}

\keywords{galaxies: evolution -- galaxies: haloes -- galaxies: statistics -- large-scale structure of Universe -- cosmology: theory}

\maketitle

\section{Introduction}

Galaxy properties depend on environment. This is reflected in the morphology--density relation, in the environmental trends of colour and star-formation activity, and in marked clustering measurements that connect galaxy populations to large-scale structure \citep{Dressler1980Morphology,Kauffmann2004Env,sheth2005marked,skibba2006marked}. A particularly suggestive example is galactic conformity: the tendency for galaxies around quenched or passive primaries to themselves be more quenched than galaxies around star-forming primaries \citep{Weinmann2006,Kauffmann2013Conformity}. On scales beyond the virial radius, this two-halo conformity has been widely connected to halo assembly bias, namely the dependence of halo clustering on secondary properties at fixed mass \citep{Wechsler2006,GaoWhite2007,Hearin2015Beyond,calderon2018conformity,pahwa2017conformity,LacernaPadillaPalma2025}. In that broad picture, galaxy properties respond to halo assembly, and halo assembly in turn retains memory of the large-scale environment.

In this letter we focus on
\begin{equation}
  \dfq(r) \equiv f_{\mathrm{Q}}(r|\mathrm{Q}) - f_{\mathrm{Q}}(r|\mathrm{SF}),
  \label{eq:def_dfq}
\end{equation}
the difference between the fraction of quenched neighbours 
($f_{\mathrm{Q}}$) around quenched primaries and the corresponding fraction around star-forming primaries at a distance $r$. Although this 
quantity has been widely used as a diagnostic of two-halo conformity, its statistical structure has not previously been written explicitly in correlation-function language. We show that $\dfq(r)$ can be expressed directly in terms of auto- and cross-correlations of the relevant galaxy samples, that its leading large-scale behaviour is naturally described by a response-like form proportional to the linear matter correlation function $\ximmlin(r)$, and that its amplitude cannot be explained by halo-mass bias alone. 
This motivates an interpretation in which $\dfq$ exposes the shape of linear matter clustering through a ratio-and-difference construction, while galaxy assembly bias sets the amplitude and helps suppress nonlinear residuals.

{This is the first paper in the series and focuses on the real-space statistic; the projected analogue, relevant for observational applications, is treated in an upcoming paper.} We derive the theoretical model for galactic conformity in Sect.~\ref{sec:theory}, describe the measurements in Sect.~\ref{sec:sims}, present the main results in Sect.~\ref{sec:results}, and place supporting common-shape and figure of merit (FoM) tests in Appendices~\ref{sec:common} and \ref{sec:fom_appendix}.

\section{Conformity, assembly bias, and linear matter clustering}
\label{sec:theory}

We develop an analytic model for conformity that makes its relation to
assembly bias explicit.  The same derivation shows why a
single-amplitude linear matter template is the natural large-scale
expectation, and why the exact ratio form can preserve this template to
smaller separations than what would be expected from the individual
correlation functions.

\subsection{Analytic model for conformity}
\label{sec:model}
Let $p$ denote a primary sample and let N and Q denote the full neighbour sample and its quenched subset. The quenched-neighbour fraction around primaries of type $p$ can be written as
\begin{equation}
  f_{\mathrm{Q}}(r|p)=
  \frac{\bar n_{\mathrm{Q}}\,[1+\xi_{p\mathrm{Q}}(r)]}
       {\bar n_{\mathrm{N}}\,[1+\xi_{p\mathrm{N}}(r)]},
  \label{eq:fQ_exact}
\end{equation}
where $\bar n_{\mathrm{Q}}$ and $\bar n_{\mathrm{N}}$ are the mean number densities of quenched neighbours and all neighbours. For quenched and star-forming primaries this gives
\begin{equation}
\dfq(r)=
\frac{\bar n_{\mathrm{Q}}}{\bar n_{\mathrm{N}}}
\left[
\frac{1+\xi_{\mathrm{QQ}}(r)}{1+\xi_{\mathrm{QN}}(r)}
-
\frac{1+\xi_{\mathrm{SFQ}}(r)}{1+\xi_{\mathrm{SFN}}(r)}
\right].
\label{eq:dfq_exact}
\end{equation}
Thus $\dfq$ is a definite combination of auto- and cross-correlation functions rather than an ad hoc environmental statistic.

We consider two models. First, in the bias-factorisation model, each
\(\xi_{ij}\) entering Eq.~(\ref{eq:dfq_exact}) is replaced by
\(b_i b_j\,\ximmlin(r)\) \citep[e.g.][]{Zehavi2011} and inserted into the full ratio before taking
the difference. This is not a one-parameter fit to \(\dfq\), but a
combination of biased linear templates for the ingredients of the
statistic.

Second, we use a single-template model for \(\dfq\). Defining
\(\bar f_Q=\bar n_Q/\bar n_N\), the weak-clustering limit where the correlation functions entering
Eq.~(\ref{eq:dfq_exact}) satisfy \(|\xi_{pQ}|,|\xi_{pN}|\ll1\), gives
\(f_Q(r|p)\simeq \bar f_Q[1+\xi_{pQ}(r)-\xi_{pN}(r)]\). Taking the
difference between quenched and star-forming primaries cancels the
constant term and yields
\begin{equation}
\dfq(r)\simeq
\bar f_Q\left[
\xi_{\mathrm{QQ}}-\xi_{\mathrm{QN}}
-\xi_{\mathrm{SFQ}}+\xi_{\mathrm{SFN}}
\right].
\label{eq:dfq_xi}
\end{equation}

If the large-scale factorisation
$\xi_{ij}(r)\simeq b_i\,b_j\,\ximmlin(r)$
holds, then
\begin{equation}
\dfq(r)\simeq
\bar f_Q\left(b_{\mathrm{Q}}-b_{\mathrm{SF}}\right)
\left(b_{\mathrm{Q}}-b_{\mathrm{N}}\right)\ximmlin(r).
\label{eq:dfq_factorized}
\end{equation}
This shows that a linear-response form is already the natural
large-scale expectation for $\dfq$, with a bias-factorisation amplitude
\begin{equation}
A_{\rm bias} \simeq \bar f_Q
\left(b_{\mathrm{Q}}-b_{\mathrm{SF}}\right)
\left(b_{\mathrm{Q}}-b_{\mathrm{N}}\right).
\label{eq:A_factorized}
\end{equation}
It therefore motivates our second model, a compact fitted template,
\begin{equation}
\dfq(r)=A_{\rm fit}\,\ximmlin(r),
\label{eq:linear_template}
\end{equation}
where $A_{\rm fit}$ is fitted directly to the measured conformity
statistic {rather than inferred from the individual $\xi_{ij}$}. Equation~(\ref{eq:A_factorized}) gives the corresponding
large-scale bias expectation, which we compare with the fitted amplitude
below.

The more interesting question is what sets the amplitude. A halo-mass-only
estimate of the quenched--star-forming bias contrast is
\begin{equation}
  \Delta b_{\mathrm{mass}} = b(M_{\mathrm{Q}})-b(M_{\mathrm{SF}}),
\end{equation}
with biases evaluated from median halo masses using the \citet{Tinker2010}
relation. {As a check, we also replace each sample bias by
\(\langle b(M_{\rm host})\rangle\), averaged over its full host-halo mass
distribution. This changes the mass-only bias contribution by factors of order unity,
but it remains below the total effective bias contribution.} We write
\begin{equation}
  \Delta b_{\mathrm{eff}} = \Delta b_{\mathrm{mass}} + \Delta b_{\mathrm{AB}},
  \label{eq:deltab}
\end{equation}
{where $\Delta b_{\mathrm{AB}}$ denotes galaxy assembly bias at
fixed halo mass: the part of the galaxy--halo connection not explained by halo mass and
suppressed by fixed-mass shuffling. It can reflect several secondary halo or
environmental variables. For the observables considered here, halo mass alone
is insufficient and the remaining signal is dominated by this fixed-mass
component.}

This interpretation also motivates a broader family of transformed
conformity statistics. Let $f_{X|H}(r)$ and $f_{X|L}(r)$ denote the
fractions of neighbours with property $X$ around the two primary classes
$H$ and $L$. We define
\begin{equation}
G_n(r)\equiv f_{X|H}^n(r)-f_{X|L}^n(r),
\label{eq:gn_def}
\end{equation}
so that the usual conformity statistic is recovered for $n=1$ and
$X=Q$, i.e. $G_{n=1}(r)=\Delta f_Q(r)$.

For \(n=3\),
\begin{equation}
G_{n=3}(r)=\Delta f_X(r)
\left[f_{X|H}^2+f_{X|H}f_{X|L}+f_{X|L}^2\right],
\label{eq:g3_factorized}
\end{equation}
so \(G_{n=3}\) preserves the same large-scale template as
\(\Delta f_X\), but smoothly reweights the signal by the local value of
the selected-neighbour fraction. This can improve the balance between
linearity and noise, although large \(n\) eventually emphasises noisy
excursions.  {From this point on, we refer to this case of transformed conformity as cubic conformity.}

\subsection{Beyond the weak-clustering regime}
{The weak-clustering derivation explains the origin of the large-scale
template, but it does not by itself determine how far this approximation
should extend. The exact ratio form suggests why it may remain useful
even when the individual correlations are no longer weak. For a general
selected neighbour population $X$ and primary classes $H$ and $L$, the
exact expression can be written as
\begin{align}
\frac{\Delta f_X(r)}{\bar f_X}
&=
\frac{1+\xi_{HX}(r)}{1+\xi_{HN}(r)}
-
\frac{1+\xi_{LX}(r)}{1+\xi_{LN}(r)}
\nonumber\\
&=
\left[1+\frac{\xi_{HX}(r)-\xi_{HN}(r)}
{1+\xi_{HN}(r)}\right]
-
\left[1+\frac{\xi_{LX}(r)-\xi_{LN}(r)}
{1+\xi_{LN}(r)}\right]
\nonumber\\
&=
\frac{D_H(r)}{1+\xi_{HN}(r)}
-
\frac{D_L(r)}{1+\xi_{LN}(r)},
\label{eq:deltafx_exact_D}
\end{align}
where $D_t(r)\equiv \xi_{tX}(r)-\xi_{tN}(r)$. Thus the statistic compares the
selected-neighbour excess relative to all neighbours, normalised by the total
neighbour abundance around the same primaries. {This expression can suppress
nonlinear and transition-scale clustering common to $X$ and $N$, but does not
imply that halo exclusion, baryonic effects, or other nonlinear terms are
absent.}}

{To make this explicit, we write
\begin{equation}
\xi_{ab}(r)=b_a b_b\,E_{ab}(r)\,\xi_{\rm mm}^{\rm lin}(r),
\label{eq:E_def}
\end{equation}
where $E_{ab}\rightarrow1$ on large scales and absorbs nonlinear,
transition-scale and one-halo departures from linear bias.  Substitution
into Eq.~(\ref{eq:deltafx_exact_D}) gives
\begin{equation}
\Delta f_X
=
{\cal A}_X(r)\,\xi_{\rm mm}^{\rm lin}(r),
\label{eq:deltafx_Aofr}
\end{equation}
with all departures from a single-amplitude template contained in
\begin{equation}
\frac{{\cal A}_{X}(r)}{\bar f_X}
=
\frac{
b_H\left[b_XE_{HX}-b_NE_{HN}\right]
}{
1+b_Hb_NE_{HN}\xi_{\rm mm}^{\rm lin}
}
-
\frac{
b_L\left[b_XE_{LX}-b_NE_{LN}\right]
}{
1+b_Lb_NE_{LN}\xi_{\rm mm}^{\rm lin}
}.
\label{eq:Aofr_E}
\end{equation}
In the large-scale limit,
${\cal A}_{X}\rightarrow \bar f_X(b_H-b_L)(b_X-b_N)$, recovering
Eq.~(\ref{eq:dfq_factorized}).  On smaller scales the individual
$E_{ab}$ need not be close to unity; the relevant condition is only that
their combination in ${\cal A}_{X}(r)$ varies slowly.  This is plausible
for assembly-biased splits at nearly fixed halo mass, where the one-halo
and transition-region neighbour profiles are similar, while the selected
fractions respond differently to large-scale environment.}

{The fitted
template tested below is therefore equivalent to asking whether
${\cal A}_{X}(r)$ { remains approximately constant over
the fitted range, close to its weak-clustering bias expectation
\({\cal A}_{X,\rm bias}\equiv \bar f_X(b_H-b_L)(b_X-b_N)\), so that
\begin{equation}
\Delta f_X(r)\simeq A_{\rm fit}\,\ximmlin(r)
\end{equation}
beyond the formal weak-clustering regime. }

\section{Simulation and measurements}
\label{sec:sims}

We use IllustrisTNG300-1 at $z=0$ \citep{Pillepich2018TNG,Springel2018TNG,Nelson2018TNG}, which adopts a flat $\Lambda$CDM cosmology with $\Omega_{\rm m}=0.3089$, $\Omega_{\Lambda}=0.6911$, $\Omega_{\rm b}=0.0486$, $\sigma_8=0.8159$, $n_{\rm s}=0.9667$, and $h=0.6774$ \citep{Springel2018TNG,Nelson2019TNGdata}. The linear matter correlation function is computed for this cosmology from CAMB \citep{CAMB}.

Primary galaxies are restricted to centrals. For the quenched analysis they
are split at ${\rm sSFR}=10^{-11}\,{\rm yr}^{-1}$; for colour and
concentration we split at the median central-galaxy value. Tracers include all
galaxies with $M_\star>10^9\,h^{-1}M_\odot$, irrespective of central or
satellite status; primaries use the same cut unless otherwise stated.
{In Appendix~\ref{sec:fom_appendix}, the central-primary number density
is varied by cumulative stellar-mass thresholds before splitting by sSFR,
colour, or concentration.} Measurements are performed in real space over
$1$--$40\,h^{-1}\,\mathrm{Mpc}$ using \textsc{Corrfunc}
\citep{SinhaGarrison2020Corrfunc}. Effective biases are measured from the same
correlations, except for the halo-mass-only estimate, where we use host masses
and the \citet{Tinker2010} relation.

\section{Results}
\label{sec:results}
The radial behaviour of the quenched fractions and of their difference reveals where the approximately linear conformity signal emerges and how it is isolated from the more nonlinear structure present in the individual terms. 
Figure~\ref{fig:main} shows the main result. The individual quenched fractions
around quenched and star-forming primaries are not single-amplitude linear
responses, but their difference is much simpler. {As anticipated by
Eq.~\ref{eq:deltafx_exact_D}, part of the nonlinear structure common to the
selected and total neighbour profiles is suppressed, leaving
\({\cal A}_Q(r)\) nearly constant over the fitted range.} Thus \(\dfq(r)\) is
well described by \(A_{\rm fit}\ximmlin(r)\) down to small separations and
remains non-zero out to \(\sim40\,h^{-1}{\rm Mpc}\), despite its amplitude of
order \(10^{-3}\).

This behaviour supports an assembly-bias interpretation of the amplitude
together with a linear-response interpretation of the radial dependence:
 the contrast between the primary samples sets the strength of the signal,
while the ratio-and-difference construction removes much of the nonlinear
structure common to the {neighbour profiles around the different primaries.} 

\begin{figure}
  \centering
  \includegraphics[width=0.94\columnwidth]{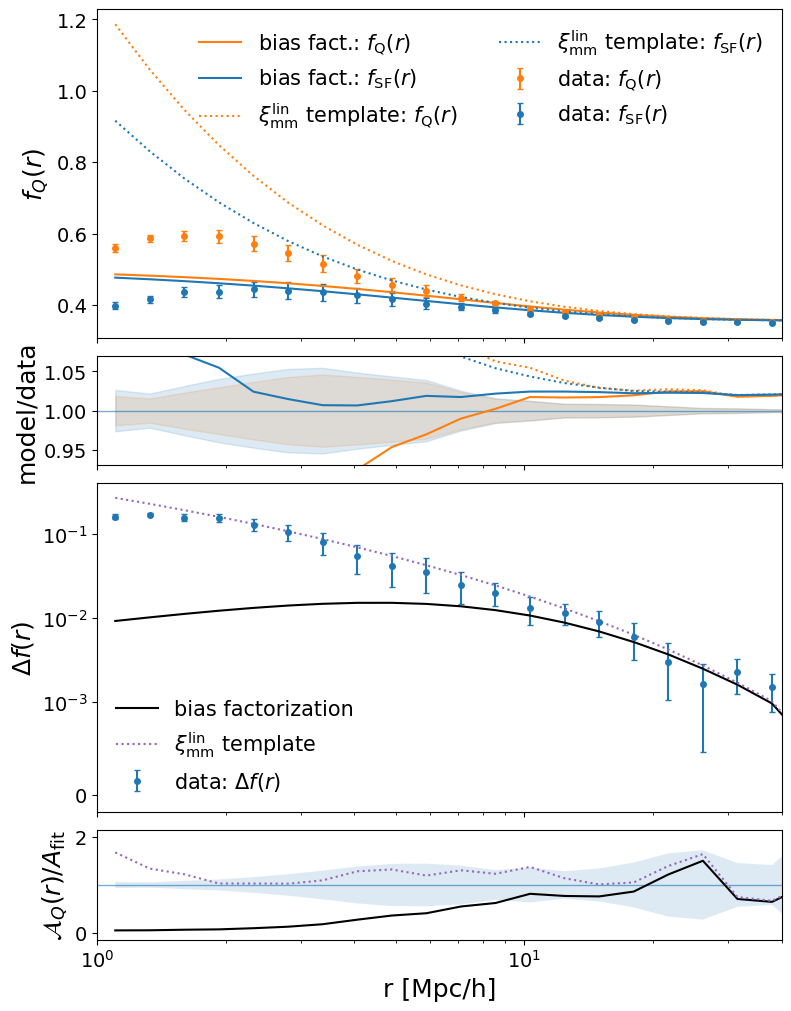}
  \caption{Fractions of quenched galaxies and galactic conformity. Top:
quenched-neighbour fractions around quenched and star-forming central
primaries, together with the corresponding bias-factorisation and linear
template predictions; colours identify the primary samples and line
styles identify the models. The bottom subpanel shows ratios between
models and data; error bars are obtained from 64 jackknife regions.
Bottom: the corresponding conformity statistic $\dfq(r)$ compared with
the best-fitting $A_{\rm fit}\ximmlin(r)$ template and with the bias-factorisation
model. Bottom subpanel: effective response normalised by the fitted amplitude,
\(\Delta f_Q(r)/[A_{\rm fit}\xi_{\rm mm}^{\rm lin}(r)]\);
unity marks the fitted linear-template response. The black solid curve
shows the bias-factorisation prediction normalised by the same fitted
template}.
  \label{fig:main}
\end{figure}

\begin{figure}
  \centering
  \includegraphics[width=0.94\columnwidth]{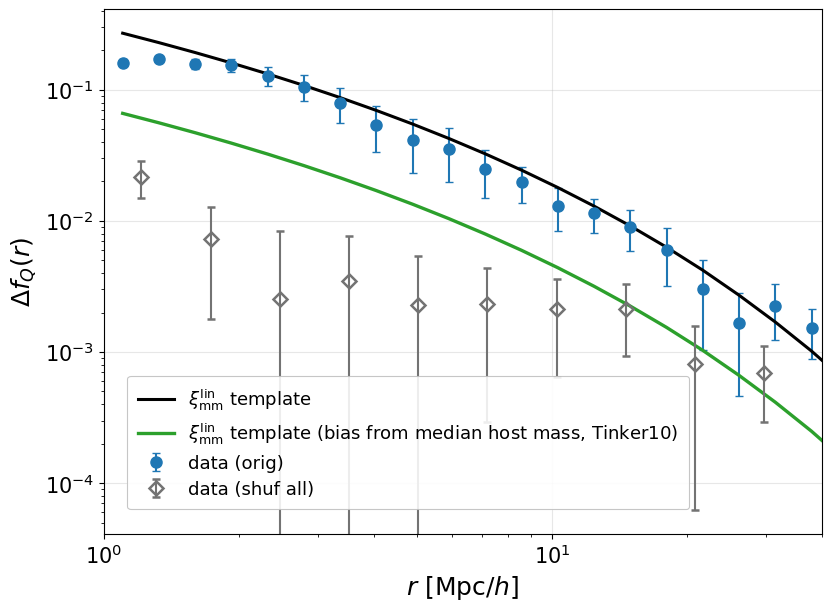}
  \caption{Shuffling test. The original conformity signal (solid symbols) is strongly suppressed in the sample obtained by shuffling galaxies between haloes in $0.1$ dex  mass bins (open squares). This demonstrates that the amplitude of $\dfq$ is dominated by {galaxy} assembly bias rather than by halo-mass selection (see the legend for description of different model lines).}
  \label{fig:shuffle}
\end{figure}

Equation~(\ref{eq:deltab}) shows that a large conformity amplitude requires a large effective bias contrast; if only the halo-mass term contributes, the predicted signal should be much smaller for typical quantities used in galactic conformity.

Figure~\ref{fig:shuffle} tests the amplitude interpretation. The mass-only
prediction uses Eq.~(\ref{eq:deltab}) with biases inferred from host
halo masses using \citet{Tinker2010}. The open symbols show catalogues in which galaxies are shuffled
among haloes in $0.1$ dex mass bins, preserving halo-mass distributions but
erasing fixed-mass correlations with secondary variables. {Both the
mass-only prediction and the shuffled result have smaller amplitudes than the
original signal. Thus halo mass alone is insufficient, and in TNG300 the
amplitude is dominated by galaxy assembly bias at fixed halo mass. The
baryonic model sets the galaxy--halo connection; shuffling isolates the part
correlated with secondary halo or environmental variables at fixed mass.}\footnote{{We also explored alternative shuffling schemes affecting only centrals, or only satellites. All shuffles suppress the original signal, confirming that the large amplitude of $\dfq$ is not tied to a particular implementation of the shuffle test. The strongest suppression occurs when central properties, or all galaxy properties, are shuffled, whereas shuffling satellites alone leaves a larger residual signal consistent with results presented by \cite{Hearin2015Beyond}.}}

The interpretation of $\dfq$ as a linear response-like statistic is further
sharpened in Appendix~\ref{sec:common}, where conformity-like statistics
based on colour and concentration are shown to share the same rescaled
large-scale shape, whereas stellar-mass and halo-mass splits show less
clean agreement with $\ximmlin(r)$.
The transformed-statistic and information-content tests are discussed in
Appendix~\ref{sec:fom_appendix}, since they support rather than define the
main claim.

\section{Discussion and conclusions}
\label{sec:conclusions}

The main result of this work is that two-halo conformity separates naturally
into a large-scale response-like component and a galaxy-assembly-bias amplitude. The
statistic $\dfq(r)$ is well described by $A_{\rm fit}\,\ximmlin(r)$ over
$\sim 2$--$40\,h^{-1}\,\mathrm{Mpc}$, {indicating that the
ratio-and-difference construction described in Sec. \ref{sec:theory} suppresses part of the nonlinear and
baryonic scale dependence present in the individual correlations.}

This interpretation is reinforced by the fact that conformity statistics based
on quenching, colour, and concentration share the same large-scale radial
dependence after amplitude rescaling, and that {fixed-mass} shuffling
{suppresses their amplitudes}. {The concentration case links this
behaviour to a standard secondary halo property associated with assembly bias,
while quenching and colour show how such fixed-mass halo dependence is mapped
into galaxy properties.} By contrast, analogous statistics built from
stellar-mass and halo-mass splits do not show comparably good agreement with
$\ximmlin(r)$. These are best viewed as stress tests rather than pure
mass-bias controls, since such splits induce much stronger abundance and
occupation differences. Their behaviour nevertheless shows that near-linearity
is not a generic consequence of taking differences of neighbour fractions. A
natural interpretation is that observables dominated by
{galaxy assembly bias} are more likely to preserve a
common large-scale template to smaller separations because they compare
populations at more nearly fixed halo mass, and therefore with more similar
virial radii and less disparate transition-region contributions. This is also
consistent with previous work showing that galaxy assembly bias can be
relatively weak near transition scales \citep{Contreras2019}. This suggests
that \(\dfq\)-like statistics can retain sensitivity to a large-scale response
down to scales that are usually regarded as nonlinear.

The transformed family $G_n$, {including the cubic conformity statistic \(G_{n=3}\) } shows that $\dfq$ is not a special one-off
construction. In Appendix \ref{sec:fom_appendix} we show that the factorised form of \(G_{n=3}\) preserves the same
large-scale template as \(\Delta f_X\), while smoothly reweighting the
signal by the selected-neighbour fractions. Intermediate transformations, particularly {the cubic conformity statistic} \(n=3\), can
therefore improve the balance between linearity depth and noise, although
the gain is limited and eventually reversed for larger \(n\). The
\(\Omega_m\) variation test in Appendix~\ref{sec:common} should likewise
be understood as a consistency check on the linear-template
interpretation, not as a competitive cosmological constraint.

Our approach fits in the spirit of transformed two-point statistics designed to recover a simpler response from non-linear fields \citep{MassaraEtAl2021Marked,Neyrinck2009LogTransform,WangEtAl2011LogDensity,PaillasEtAl2022DensitySplit}, although the specific ratio-and-difference conformity statistic $\dfq$ and its transformed family $G_n$ are introduced here.

{Overall, \(\dfq\) is a compact real-space conformity statistic in which
galaxy assembly bias controls the amplitude, while the
radial dependence traces the linear matter clustering template
\(\ximmlin(r)\).}

\begin{acknowledgements}
We thank the IllustrisTNG team for making their simulation data publicly available. We are grateful to Sergio Contreras, Lucia Perez and Idit Zehavi for useful discussions on assembly bias, conformity, and large-scale biasing. NP acknowledges support from PICT Raices Federal 2023-0002. IL acknowledges support from the ANID FONDECYT Regular grant 1261197.
\end{acknowledgements}

\bibliographystyle{aa}
\bibliography{PadillaN}

\begin{appendix}

\section{Common-shape and cosmology-shape tests}
\label{sec:common}
{Figure~\ref{fig:common} sharpens the interpretation of $\dfq$ as a linear
response-like statistic. The solid curves show conformity statistics based on
quenching, colour, and concentration. We have checked that in all three cases, fixed-mass shuffling
strongly suppresses the amplitude, which indicates that the signal is dominated by
galaxy assembly bias rather than by halo mass alone. The concentration
case is useful because concentration is a standard secondary halo property
associated with assembly bias, providing a concrete halo-level counterpart to
the quenching and colour tests.}  These signals have similar scale dependence in the upper panel and, once normalised by the best-fitting linear template, remain close to $\xi_{mm}^{\rm lin}(r)$ in the lower panel. This indicates that quenching, colour, and concentration behave consistently with a common response-like form when the ratio-and-difference construction isolates a signal dominated by galaxy assembly bias.
The amplitudes of $\Delta f_Q$ and $\Delta f_{\rm red}$ are also of the same broad order as previously reported two-halo conformity signals in simulations and data, although detailed comparison is non-trivial because the precise amplitudes depend strongly on the choice of primary sample, central selection, projection, and observable \citep{Kauffmann2013Conformity,Hearin2015Beyond,Bray2016,Lacerna2018Conformity,Lacerna2022,Ayromlou2022PhysicalOrigin}.

The dotted curves show analogous statistics based on stellar-mass and halo-mass
splits. {For these tests we define the $H$ and $L$ samples using thresholds of
$M_{\rm halo}=10^{13}\,h^{-1}M_\odot$ and
$M_\star=10^{11}\,h^{-1}M_\odot$, respectively.} These are best viewed as
stress tests rather than pure mass-bias controls, since such splits induce much
stronger abundance and occupation differences than the conformity observables.
Their poorer agreement with the linear template therefore shows that
near-linearity is not a generic property of all $\Delta f$ constructions.
{The reason is that stellar-mass and halo-mass splits directly compare
populations with different halo masses, satellite fractions, virial radii, and
transition-scale neighbour profiles. They therefore leave larger residual
nonlinear structure in the ratio-and-difference statistic. By contrast,
quenching, colour, and concentration are more closely linked to assembly bias at nearly fixed halo mass, so they compare populations with more similar
halo-mass and transition-scale structure.}

\begin{figure}
  \centering
  \includegraphics[width=0.98\columnwidth]{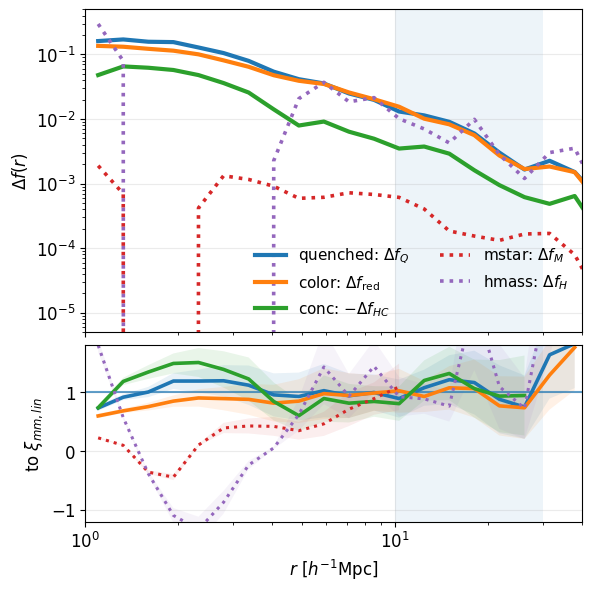}
  \caption{Common-shape test. Top: conformity-like observables for
quenching, colour, concentration, stellar-mass, and halo-mass splits.
Bottom: the same signals divided by their fitted amplitudes and by
\(\ximmlin(r)\). Quenching, colour, and concentration show a common
rescaled radial dependence, while stellar-mass and halo-mass splits
deviate more strongly, especially at small separations.  The shaded area shows the range of scales over which $A_{\rm fit}$ is obtained. }
  \label{fig:common}
\end{figure}
\begin{figure}
  \centering
  \includegraphics[width=0.94\columnwidth]{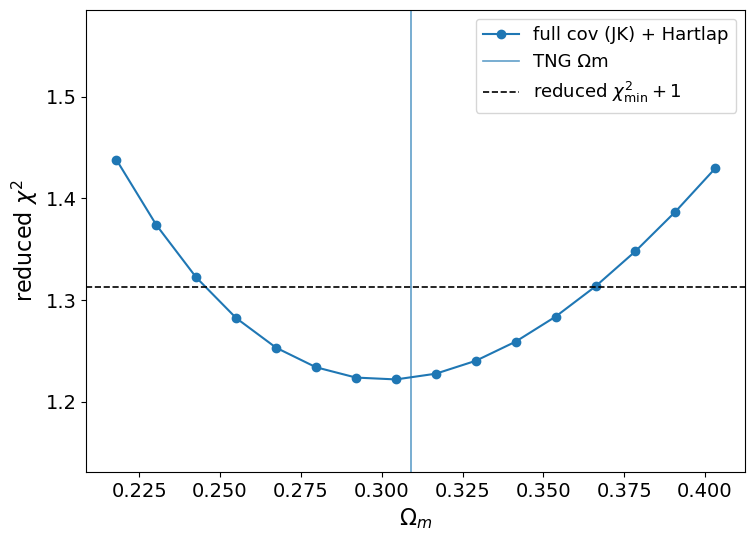}
  \caption{Shape-only $\Omega_m$ fit using $\dfq(r)$ over $3$--$40\,h^{-1}\,\mathrm{Mpc}$, with the amplitude marginalised. The vertical line shows the TNG300 cosmology.}
  \label{fig:omegam}
\end{figure}

A further test of the linear-template interpretation is to vary the shape of $\xi_{mm}^{\rm lin}(r)$ through $\Omega_m$, while marginalising over the overall amplitude, and ask whether the value preferred by $\dfq(r)$ is the one corresponding to the TNG300 cosmology. Figure~\ref{fig:omegam} shows the result using all galaxies with $M_\star > 10^9\, h^{-1} M_\odot$ and separations between $3$ and $40\,h^{-1}\,\mathrm{Mpc}$. The best-fitting value is consistent with the input cosmology of the simulation. The point of this exercise is not to claim $\dfq$ as a competitive cosmological probe, but to show that once its amplitude is left free, its measured radial dependence is accurately described by the $\Omega_m$-dependent shape of the linear matter correlation function.

\section{Linearity depth and information content}
\label{sec:fom_appendix}

\begin{figure*}
  \centering
  \includegraphics[width=0.98\columnwidth]{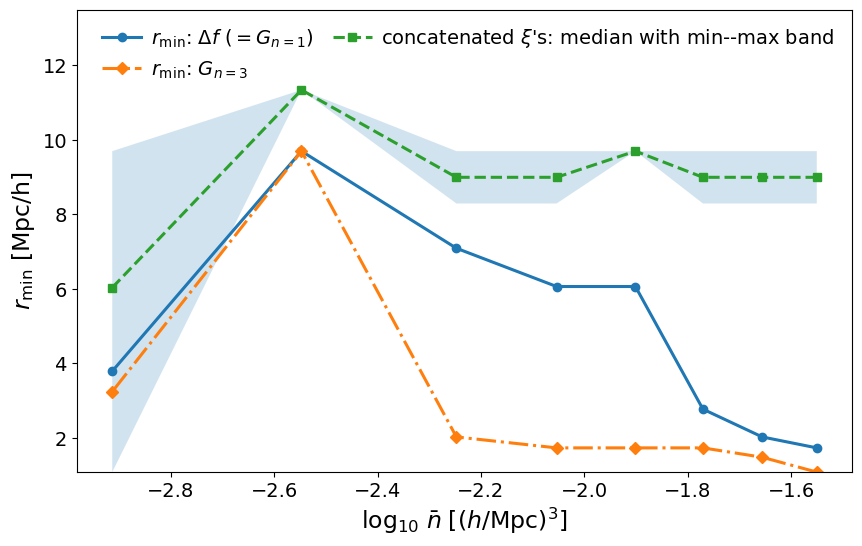}
  \includegraphics[width=0.98\columnwidth]{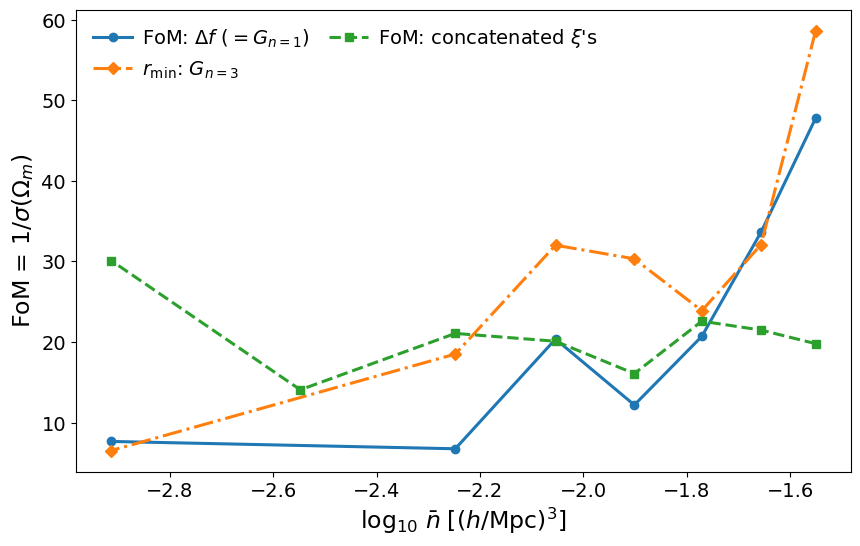}
  \caption{Linearity depth and information content. Left: minimum separation
\(r_{\min}\) down to which the shape fit to \(\ximmlin(r)\) remains
consistent with the TNG300 cosmology as a function of number density, for \(\dfq\) (blue solid line), \(G_{n=3}\) (orange dot-dashed line), and the
concatenated correlation-function data vectors entering their
construction (green dashed line). {Here, the number density is that of the central-primary sample after applying
a cumulative stellar-mass threshold; at each threshold the selected centrals are
then split by sSFR, colour, or concentration.} The shaded band spans the range of \(r_{\min}\) values
among these statistics. Right: corresponding FoM for the same
observables. The \(n=3\) transformation provides a modest improvement
over \(\Delta f_Q\) in some regimes, while the concatenated \(\xi\)
vectors recover more information in others.}
  \label{fig:fom}
\end{figure*}
\begin{figure}
  \centering
  \includegraphics[width=0.98\columnwidth]{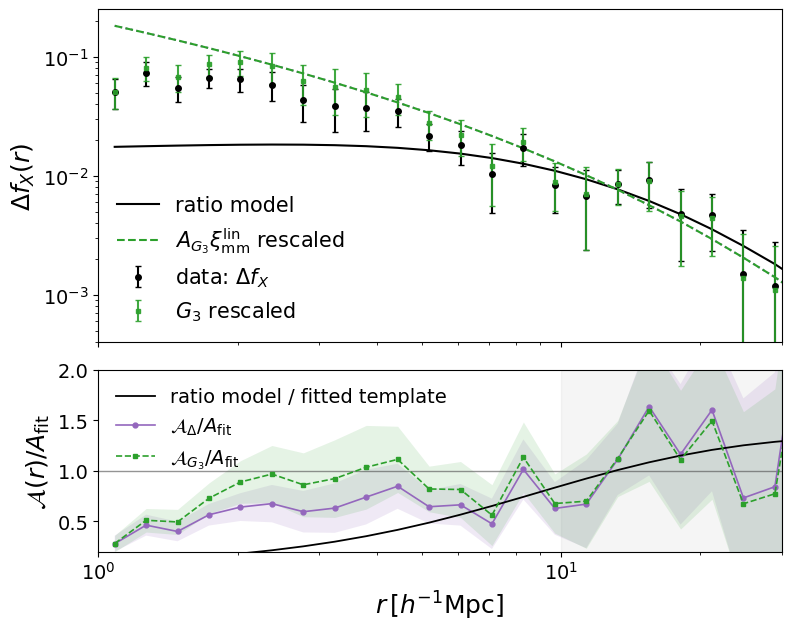}
  \caption{
  Example of the $G_{n=3}$ transformation for a low-number-density
stellar-mass-threshold sample,
$\log_{10}(M_\star/[h^{-1}M_\odot])>10.25$.
Top: measured $\Delta f_X(r)$ compared with the ratio model and the
best-fitting linear template, $A_\Delta \xi_{\rm mm}^{\rm lin}$.
The $G_{n=3}$ measurement is rescaled by $A_\Delta/A_{G_3}$ to place it
on the $\Delta f_X$ vertical scale.  With this normalisation, the
best-fitting $G_{n=3}$ template is identical to
$A_\Delta \xi_{\rm mm}^{\rm lin}$ and is not shown separately.
Bottom: normalised response functions for $\Delta f_X$, the ratio
model, and $G_{n=3}$, each divided by its own best-fitting linear
template.  The cubic conformity statistic produces a flatter response
over a wider radial range, explaining why it can remain consistent with
the linear matter template down to smaller separations than the
$n=1$ statistic in Fig.~\ref{fig:fom}.
  }
  \label{fig:gn3_example}
\end{figure}

{Figure~\ref{fig:fom} places $\dfq$ within the broader family of
transformed conformity statistics defined in Sect.~\ref{sec:model},
$G_n=f_{X|H}^n-f_{X|L}^n$.  In particular, the factorised form in
Eq.~(\ref{eq:g3_factorized}) shows that $G_{n=3}$ is a smooth
reweighting of $\Delta f_X$, so the tests below should be interpreted as
measuring whether this reweighting improves the linearity-depth/noise
trade-off, rather than as a new independent conformity signal.  The
figure separates two distinct questions: how far into small scales the
linear-template description remains valid, and how much statistical
information the observable actually carries.}

To estimate the former, we determine for each statistic the minimum fitted separation $r_{\min}$ for which the $\Omega_m$ value recovered from the $\ximmlin$ fit remains consistent with the TNG300 cosmology. We do this both for the transformed conformity statistic itself, $\gn$, and for the concatenated set of correlation functions entering its ratio-model construction, in order to test which of the two retains linear-template behaviour to smaller scales and whether the corresponding information content is comparable. To do this we estimate covariance matrices from 64 jackknife regions for both $\gn$ and the corresponding concatenated $\xi$ data vectors, and use them to compute a figure of merit (FoM) for the shape fit. {The jackknife covariances are internal error estimates and do not include the full
cosmic variance of independent TNG300 volumes; however, $\dfq$,
$G_{n=3}$, and the concatenated $\xi$ vectors are all treated with the same
covariance scheme, so the relative comparison is meaningful even if the
absolute FoM values should not be over-interpreted.}

The left panel shows the resulting $r_{\min}$ values. Here the clearest winner is {the cubic conformity statistic} $n=3$, which reaches smaller fitting scales than the binary $n=1$ case and also smaller scales than the concatenated $\xi$ description over much of the number-density range. This indicates that the transformed conformity statistic can isolate a component whose shape remains close to $\ximmlin$ deeper into the nonlinear regime than either $\dfq$ itself or the full set of ingredients from which it is built. 

The right panel shows the corresponding FoM comparison, where the picture is more nuanced. At high space densities the $n=3$ statistic performs somewhat better, supporting the idea that $\dfq$ is not merely a lucky construction but the first member of a broader family of observables that project the same assembly field with different efficiency. At lower space densities the concatenated $\xi$ measurements recover the advantage, as expected once compression into a single statistic begins to lose information in the shot-noise-dominated regime. Even so, the broadly similar FoM values are themselves notable: they show that $\dfq$ captures a substantial fraction of the relevant real-space information in a much more compact observable.  The figure therefore suggests a genuine but limited optimisation: the intermediate case $n=3$ (cf. Section \ref{sec:model}) improves the balance between linearity and noise relative to $n=1$. Tests with larger exponents, not shown here, indicate that the gain is not monotonic and is eventually lost as the statistics become noisier.

Figure~\ref{fig:gn3_example} illustrates this behaviour {of the cubic conformity statistic} for one of the
low-space-density samples, using the threshold
$\log_{10}(M_\star/[h^{-1}M_\odot])>10.25$.  In this regime the usual
$\Delta f_X$ statistic already follows the linear matter template on
large scales, but its normalised response departs from a constant at quite large
 separations $\sim8~h^{-1}$Mpc.  The {cubic conformity statistic}, $G_{n=3}$, rescaled to the same
large-scale amplitude, shows a flatter response over a wider radial
range.  This example makes explicit why the $n=3$ statistic can reach a
smaller acceptable $r_{\min}$ in Fig.~\ref{fig:fom}: the cubic
transformation does not introduce a new large-scale shape, but changes
the radial weighting of the same conformity signal in a way that partly
suppresses the residual small-scale curvature.

\end{appendix}
\end{document}